# Optical frequency combs generated mechanically

M. Sumetsky

Aston Institute of Photonic Technologies, Aston University, Birmingham B4 7ET, UK

m.sumetsky@aston.ac.uk

It is shown that a highly equidistant optical frequency comb can be generated by the parametric excitation of an optical bottle microresonator with nanoscale effective radius variation by its natural mechanical vibrations.

## 1. Introduction

An optical spectrum, which possesses equidistant and coherent frequency resonances, is commonly named an optical frequency comb (OFC). Highly equidistant OFCs can be generated by electro-optical modulation of laser cavities [1] and microresonators [2]. Alternatively, broadband and exceptionally equidistant OFC can be generated with mode-locked lasers [3-6] and by nonlinear parametric excitation of modes in high Q-factor optical microresonators [7-12].

In addition to the previously developed approaches, here we show that highly equidistant and moderately broadband combs can be generated by parametric excitation of an optical mode of a bottle resonator by its natural mechanical vibrations. It is well known that mechanical vibrations of optical microresonators can be exceptionally monochromatic (like, e.g., those in quartz watches [13, 14]) and possess a remarkably high Q-factor [15, 16]. Usually, mechanical oscillations, being very small in magnitude, only slightly affect the spectrum of optical resonator by perturbing its physical dimensions and refractive index. In the first order of perturbation theory, mechanical oscillations with frequency $f$ modify optical eigenfrequencies $\nu_m$ of the resonator by splitting off the resonances at $\nu_m \pm f$. The latter can be observed in the optical spectrum as very weak peaks. However, the situation can radically change if the series of eigenfrequencies $\{\nu_m\}$ is equidistant, i.e., $\nu_{m+1} - \nu_m = \Delta\nu_0 = const$. Then, oscillations with mechanical frequency $f$ close to $\Delta\nu_0$ can give rise to the parametric excitation of optical modes with eigenfrequencies $\{\nu_m\}$ and creation of an optical frequency comb. The effect is similar to that observed in electro-optical modulation of whispering gallery mode (WGM) microresonators with quadratic optical nonlinearity [2]. However, it is based on natural rather than externally modulated excitations of optical modes and is not associated with special electro-optical properties of the microresonator material.

In practice, the eigenfrequencies $\{\nu_m\}$ cannot be perfectly equidistant and it should be assumed that $\nu_{m+1} - \nu_m = f + \delta\nu_m$, where $|\Delta\nu_m| \ll f$. In this paper, we show that an elongated optical bottle microresonator with nanoscale variation of effective radius can have an equidistant spectrum which matches its natural mechanical vibrations. We calculate the inelastic resonant transmission amplitude through parametrically excited bottle microresonator and determine the structure of generated OFC with exceptionally accurate spacing $f$. The power of OFC teeth depends on the variations $\delta\nu_m$ and can be large if the optical and mechanical Q-factors of the microresonator are high and deviations $\delta\nu_m$ are small.

## 2. A bottle microresonator with equidistant spectrum

An optical bottle microresonator is defined as a segment of an optical fiber with a bottle-shaped effective radius variation (ERV) $r(z)$ (Fig. 1) [17]. A strongly elongated bottle microresonator with ERV

$$r(z) = r_0 \cos(\kappa z), \quad \kappa = \frac{1}{\sqrt{r_0 R_0}}, \quad -\pi < \kappa z < \pi, \tag{1}$$

possesses the eigenfrequencies determined from the semiclassical quantization rule as [17, 18]:

$$\nu^{\pm}_{m,p,q} = \mu_0 + \chi_0^{\pm} + m\Delta\nu_{az} + q\Delta\nu_{ax} \tag{2}$$

$$\Delta\nu_{az} = \frac{c(1 + 2^{-1/3}\zeta_p m^{-2/3})}{2\pi n_0 r_0}, \quad \Delta\nu_{ax} = \frac{c}{2\pi n_0 \sqrt{r_0 R_0}}. \tag{3}$$

In Eqs. (1), (2) and (3), $r_0$ and $R_0 \gg r_0$ are the azimuthal and axial radii of the resonator at its center $z = 0$ shown in Fig. 1, $\chi_0^{\pm}$ is the polarization dependent shift, $c$ is the speed of light, $n_0$ is the refraction index of the bottle resonator material, $\zeta_{1,2,3,...} = 2.338, 4.088, 5.521...$ are roots of the Airy function, and integers $m, q$, and $p$ are azimuthal, radial, and axial quantum numbers, respectively. Since the spacing of the axial eigenfrequency series, $\Delta\nu_{ax}$, in Eq. (3) does not depend on quantum numbers, these series are dispersionless. For the characteristic azimuthal radius $r_0 \sim 100$ µm, the realistic values of $\Delta\nu_{ax}$ found from Eq. (2) range from $\sim 100$ GHz for $R_0 \sim 1$ mm to $\sim 100$ MHz for $R_0 \sim 1$ km. Remarkably, the gigantic 1 km axial radius $R_0$ is possible to introduce experimentally without increasing the microresonator dimensions (e.g., the parabolic bottle microresonator delay line experimentally demonstrated in [19] had $R_0 = 1.4$ km).

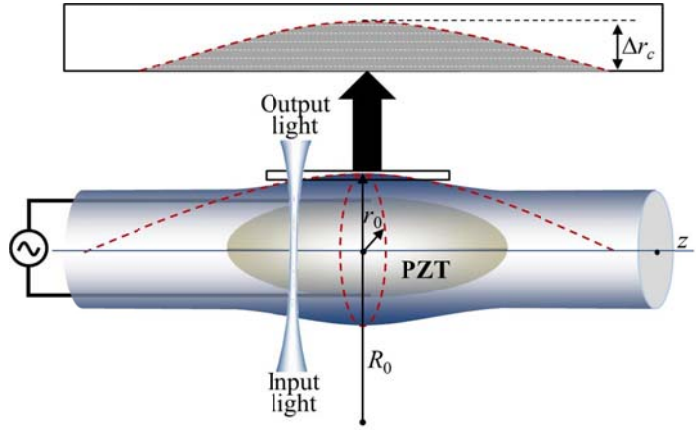

Fig. 1. Illustration of a bottle microresonator internally driven by an imbedded piezoelectric actuator (PZT). Brown dashed curve shows the ERV defined by Eq. (1). Inset: magnified profile of ERV with nanoscale height $\Delta r_c$.

## 3. Feasibility of optical frequency comb generation by natural vibrations of a bottle resonator

The idea of mechanical generation of optical frequency combs in a bottle resonator with close to equidistant spectrum is clarified in Fig. 1 and 2. As illustrated in Fig. 1, mechanical oscillations with natural frequency $f$ can be excited by internal piezoelectric actuator imbedded into the core of microresonator [20] and driven by an electric pulse modulated with the same frequency $f$ (Fig. 1). The pulse is adiabatically turned off during the time much shorter than the oscillation lifetime so that the observed comb corresponds to natural vibration frequency of microresonator. The frequency of input light, $v_1$, is tuned to the resonance frequency $v_q$ of microresonator (for brevity, here and below we omit the fixed quantum numbers $m$ and $p$). The effect of mechanical vibration with frequency $f$ consists in parametric excitation of WGMs with eigenfrequencies $v_{q+m}$ which are separated from the input frequency $v_1 = v_q$ by $mf$ (Fig. 2).

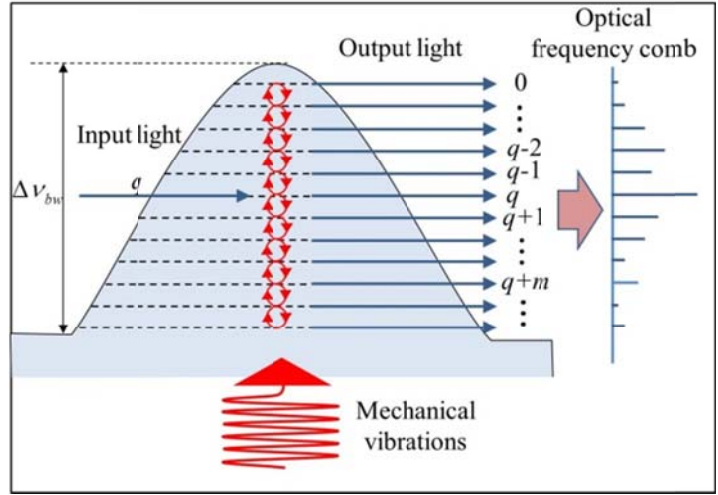

Fig. 2. Illustration of OFC generation by mechanical excitation of a bottle microresonator.

The natural frequencies of mechanical vibrations are determined by the shape and mechanical properties of microresonator. For microresonator radius $r_0 \sim 20\,\mu m$, the characteristic voltage required to introduce vibrations with amplitude 1 nm is ~10 V [20]. Natural frequencies of vibrations can be estimated as $f \sim v_s / (2\pi \lambda_s)$, where $v_s$ and $\lambda_s$ are the speed and wavelength of sound in the microresonator material. The wavelength of radial modes is $\lambda_s \leq r_0$, which gives $f \geq 100$ MHz for a bottle resonator with radius $r_0 \sim 20\,\mu m$ and $v_s \sim 5$ km/s. Equalizing $\Delta v_{ax}$ from Eq. (3) and $f = v_s / (2\pi \lambda_s)$ we have:

$$R_0 \sim \left(\frac{c\lambda_s}{n_0 v_s}\right)^2 \frac{1}{r_0} \qquad (4)$$

For example, a silica bottle resonator with radius $r_0 = 20\,\mu m$ should have the axial radius $R_0 \sim 300$ m for $r_0 / \lambda_s \sim 10$. A condition similar to Eq. (4) for the frequency spacing along the azimuthal quantum number $m$ in Eqs. (2), (3) (resembling that for toroidal and ring resonators) reads $\Delta v_z = c / (2\pi n_0 r_0) \sim v_s / (2\pi \lambda_s)$. For realistic dimensions of a microresonator of the order of 1 cm or less, the latter equation yields $\lambda_s \leq 0.1\,\mu m$. The small lifetime of such mechanical vibrations ≤ 1 ns (which drops quadratically with wavelength $\lambda_s$ [21]) makes them unpractical for mechanic generation of OFC in toroidal and ring resonators.

## 4. Transmission amplitude of a bottle resonator with nanoscale radius variation

Thus, very large axial radius $R_0$ of the bottle microresonator is required to match the spacing between its optical eigenfrequencies and a natural frequency of vibrations. At the same time, the desirable small microresonator dimensions assume its small axial length $L$. Consequently, the microresonator has to have very small magnitude of ERV $\Delta r_c \sim L^2/(2R_0)$. For example, for $R_0 \sim 1$ km and $L \sim 5$ mm this magnitude is $\Delta r_c \sim 10$ nm, characteristic for microresonators fabricated in SNAP technology [19, 22]. In spite of so small nanoscale height, this resonator supports a large number of axial modes. Here we will consider only a single series of modes (i.e., fixed $m$ in Eq. (2)) with eigenfrequencies localized near frequency $\nu_0$. For characteristic $\nu_0 \sim 400$ THz and fiber radius $r_0 \sim 20$ μm, the bandwidth of this series is $\Delta \nu_{bw} = \nu_0 \Delta r_c / r_0 \sim 200$ GHz. It follows from Eq. (3) that the spacing between eigenfrequencies of this series is $\Delta \nu_{ax} \sim 200$ MHz and, thus, it includes $\sim 1000$ elements.

Here we consider transitions between modes of such series generated by mechanical vibrations of frequency $f$ which is close to the eigenfrequency spacing $\Delta \nu_{ax}$. More precisely, we consider an optical bottle microresonator, which has a series of eigenfrequencies $\nu_q$ with almost equidistant separation, i.e., $\nu_{q+1} - \nu_q = \Delta \nu_{ax}^{(0)} + \delta \nu_q$, where $|\delta \nu_q| << \Delta \nu_{ax}^{(0)}$. The time-dependent resonance amplitude of a WGM propagating along the optical fiber with nanoscale radius variation is described by the Schrödinger equation [23]:

$$\frac{i}{2\pi\nu_0}\frac{\partial \Psi}{\partial t} = -\frac{1}{2k_0^2}\frac{\partial^2 \Psi}{\partial z^2} + V(z,t)\Psi, \quad V(z,t) = -\frac{\Delta r(z,t)}{r_0}, \quad k_0 = \frac{2\pi n_0 \nu_0}{c}. \tag{5}$$

We assume that an eigenfrequency of mechanical oscillation of this microresonator, $f$, is close to the separation of optical eigenfrequencies by setting $f = \Delta \nu_{ax}^{(0)}$. In the presence of this oscillation, the behavior of the resonator is described by Floquet quasi-states [24]

$$\Psi_q(z,t) = \exp(-2\pi\zeta_q t)\Phi_q(z,t), \tag{6}$$

where $\zeta_q$ is a quasi-frequency and $\Phi_q(\mathbf{r},t)$ is a periodic function of time with period $T = 1/\operatorname{Re} f$:

$$\Phi_q(z,t) = \sum_m \exp\left[-2\pi(im\operatorname{Re} f + |m|\operatorname{Im} f)t\right]U_{q,m}(z). \tag{7}$$

In this equation, we introduced $\operatorname{Im} f$ which characterized the attenuation and Q-factor $Q_{mech} = \operatorname{Im} f / \operatorname{Re} f$ of mechanical oscillations. Due to the proximity of frequency $f$ and spacing between optical eigenfrequencies, $\nu_{q+1} - \nu_q$, we will assume that the transition amplitudes between quasi-states $\Psi_q(\mathbf{r},t)$ is much greater than transitions between these quasi-states and other quasi-states of the resonator. Then, the expression for the non-stationary Green's function of the resonator [25] will include only the series $\{\Psi_m(\mathbf{r},t)\}$ of our interest and can be written as

$$G(z_1,z_2,t_1,t_2) = \theta(t_1-t_2)\sum_q \Psi_q(z_1,t_1)\Psi_q^*(z_2,t_2)$$
$$= \theta(t_1-t_2)\sum_{q,m_1,m_2}\exp\left[-2\pi i\zeta_q(t_1-t_2) - 2\pi i(m_1 t_1 - m_2 t_2)\operatorname{Re} f - 2\pi(|m_1|t_1 + m_2|t_2|)\operatorname{Im} f\right]U_{q,m_1}(z_1)U_{q,m_2}^*(z_2). \tag{8}$$

The eigenfrequencies of a high Q-factor optical microresonator are usually detected by measuring the resonant transmission amplitude of light evanescently coupled to the resonator though a waveguide (Fig. 1). For a resonant WGM which circulates and slowly propagates along the bottle resonator axis $z$, the propagation constant $\beta(z)$ is small, $\beta(z) << k_0$, and, therefore, the characteristic variation length $\Lambda$ of WGM along axis $z$ is large compared to the wavelength of radiation, $\Lambda \gg \lambda_0 = c/(2\pi\nu_0)$. Typically, the width of the coupling waveguide $d$ is comparable with the radiation wavelength so that $d \sim \lambda_0 \ll \Lambda$. Under these conditions, the input light with frequency $\nu_1$ can be introduced by adding the following source term to the right hand side of Eq. (5):

$$A_{in}(z,t,\nu_1) = A_0 \delta(z-z_0) \exp(-2\pi i \nu_1 t) \times \begin{cases} \exp(2\pi\gamma t) & t<0 \\ 1 & t \geq 0 \end{cases} \quad (9)$$

where $\delta(z)$ is the delta-function and $\gamma^{-1}$ characterizes the switching time, which is usually much greater than the life time of eigenstates, i.e., $|\text{Im}(\zeta_m)| \ll \gamma$. Then, for weak coupling between the input-output waveguide and resonator (strongly under-coupling regime [26]), the inelastic output amplitude ($|m_2|+|m_1| \neq 0$) is found from Eqs. (8) and (9) as

$$A_{out}(\nu_1,\nu_2) = \int_{-\infty}^{\infty} dt_2 \int_{-\infty}^{t_2} dt_1 \int_{-\infty}^{\infty} dz_1 G(z_1,z_2,t_1,t_2) A_{in}(z_1,t_1,\nu_1) = \\ = A_0 \sum_{q,m_1,m_2} \frac{U_{q,m_1}(z_0) U^*_{q,m_2}(z_0)}{[(m_2-m_1)\text{Re}\,f - i(|m_2|+|m_1|)\text{Im}\,f + \nu_2 - \nu_1](\zeta_q + m_2 \text{Re}\,f - i|m_2|\text{Im}\,f - \nu_1)}, \quad (10)$$

Selecting the sequence of resonance terms in Eq. (10) with varying $m_1$ and fixed $q$ and $m_2$ and setting the input radiation frequency $\nu_1$ equal to the minimum value of denominator in this equation,

$$\nu_1 = \text{Re}(\zeta_q) + m_2 \text{Re}\,f, \quad (11)$$

we determine the equidistant comb in the output spectrum with teeth at

$$\nu_2 = \text{Re}(\zeta_q) + m_1 \text{Re}\,f, \quad m_1 = 0, \pm 1, \pm 2, \ldots \quad (12)$$

where the range of $m_1$ is limited by the bandwidth of the resonator spectrum (Fig. 2). Eqs. (11) and (12) are symmetric with respect to substitution $1 \leftrightarrow 2$. However, due to the asymmetry of Eq. (10) with respect to this substitution, fixing the output frequency $\nu_2$ and varying the input frequency $\nu_1$ will not generate the comb of similar strength. Eqs. (10)-(12) show that there exist equally spaced combs of the transmission amplitude corresponding to $\{\text{Re}(\zeta_q) + m\text{Re}\,f\}$ with fixed $q$ and variable $m$. These combs are analyzed below for the special case of a bottle microresonator with time dependent quadratic radius variation.

The power of the comb teeth defined by Eqs. (11) and (12) is strongest when the input frequency is equal to the resonance, (i.e., $m_2 = 0$). In this case

$$\begin{aligned} \nu_1 &= \text{Re}(\zeta_q), \\ \nu_2 &= \text{Re}(\zeta_q) + m \text{Re}\,f, \quad m = \pm 1, \pm 2, \ldots \end{aligned} \quad (13)$$

The power of teeth of this comb is determined from Eq. (10) as

$$\Omega_{2q,m} = \frac{|A_0 W_{2q,0} W^*_{2q,m}|^2}{(\text{Im}\,f \,\text{Im}\,\zeta_q)^2 m^2}. \quad (14)$$

It follows from this equation that the power of teeth is proportional to the squared product of the optical Q-factor, $Q_{opt} = \mathrm{Re}\,\nu_q / \mathrm{Im}\,\nu_q$, and mechanical Q-factor, $Q_{mech} = \mathrm{Re}\,f / \mathrm{Im}\,f$.

## 5. Parametric excitation of a parabolic bottle resonator

The magnitudes of comb teeth in Eq. (10) can be determined analytically for a bottle resonator with quadratic nanoscale ERV perturbed by harmonic oscillations:

$$\Delta r(z,t) = \frac{z^2}{2R_0}\left(1 + 2\varepsilon \sin(2\pi f t)\right), \quad \varepsilon \ll 1. \tag{15}$$

The spacing of unperturbed eigenfrequencies $\Delta \nu_{ax}$ of this resonator is determined by Eq. (3). It is assumed that the light source (Eq. (9)) is positioned in the center of resonator, i.e., $z_0 = 0$. Then, due to the spatial symmetry of Eqs. (5) and (9) with respect to $z_0 = 0$, coupling to asymmetric (odd) states of resonator are zeroed and only the even states are excited (Fig. 3). The frequency of mechanical vibrations $f$ is assumed to be close to the spacing of eigenfrequencies of these states, $2\Delta\nu_{ax}$:

$$f = 2\Delta\nu_{ax}\left(1 + \frac{\delta}{2}\right), \quad \delta \ll 1. \tag{16}$$

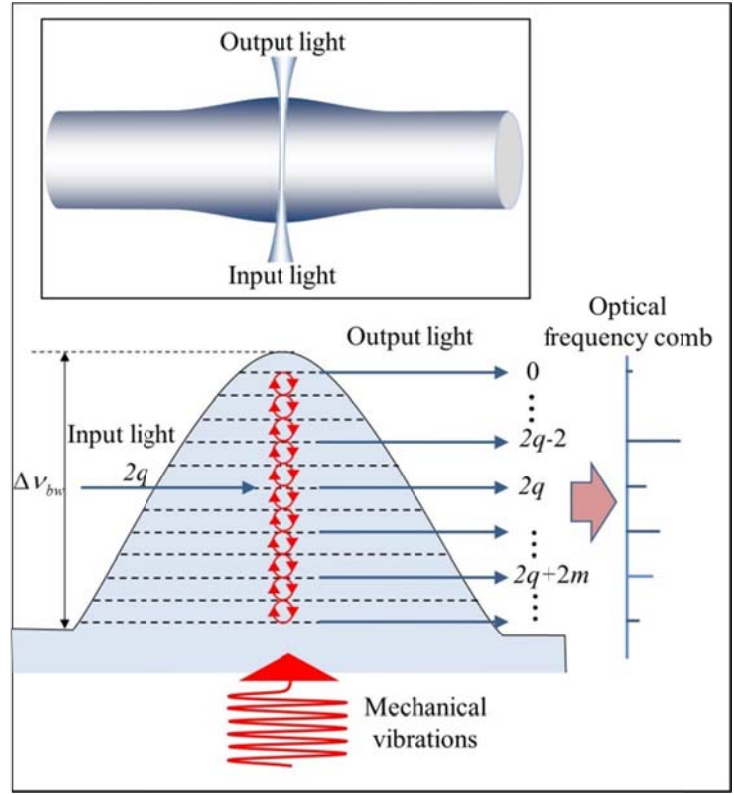

Fig. 3. Illustration of the mechanical generation of an optical frequency comb in a bottle microresonator with quadratic ERV perturbed by harmonic oscillations.

Eq. (5) with radius variation defined by Eq. (15) is a well-known Schrödinger equation with time-dependent quadratic potential that can be solved analytically [24]. The normalized solutions of this equation are:

$$\Psi_q(z,t) = \pi^{-1/4}[2^q q! y(t)]^{-1/2}\left(\frac{y^*(t)}{y(t)}\right)^{q/2} \exp\left(\frac{iy_t(t)z^2}{2y(t)}\right) H_q\left(\frac{z}{|y(t)|}\right). \tag{17}$$

Here $y(t)$ satisfies the Mathieu equation:

$$y_{tt} + (2\pi\Delta v_{ax})^2[1+2\varepsilon\sin(2\pi ft)]y = 0 \tag{18}$$

The Floquet solution of this equation found in the first order in $\delta$ and $\varepsilon$ is

$$y(t) = (\pi f)^{-\frac{1}{2}} \exp(-\pi i \alpha t)\left[\kappa^{-\frac{1}{2}}\cos(\pi ft) + i\kappa^{\frac{1}{2}}\sin(\pi ft)\right],$$

$$\alpha = \frac{f}{2}\sqrt{\delta^2 - \varepsilon^2}, \quad \kappa = \sqrt{\frac{\delta - \varepsilon}{\delta + \varepsilon}}. \tag{19}$$

In derivation of Eq. (19), it was assumed that the Floquet solution is stable, which is valid for relatively small amplitude of vibrations, $\varepsilon < \delta$. Using Eqs. (17) and (19) we expand the even solutions at $z = 0$, $\Psi_{2q}(0,t)$, into Fourier series:

$$\Psi_{2q}(0,t) = \Lambda_q \exp\left\{i\pi\left[\left(2q+\frac{1}{2}\right)(\alpha-f)\right]t\right\} \sum_{m=-\infty}^{q} C_{q,m} \exp(2\pi imft),$$

$$\Lambda_q = \frac{[(2q)!]^{\frac{1}{2}}}{(-2)^q q!}\left(\frac{\pi f \kappa}{\pi(1+\kappa)}\right)^{\frac{1}{2}}, \tag{20}$$

$$C_{q,m} = \sum_{n=\max(0,m)}^{2q} \binom{2q}{m}\binom{-q-\frac{1}{2}}{n-m}\left(\frac{1-\kappa}{1+\kappa}\right)^{2n-m}.$$

Finally, using Eqs. (10) and (20), we find for the inelastic transmission amplitude:

$$A_{out}(v_1,v_2) = -A_0 \sum_{q,m_1,m_2} \frac{W_{2q,m_1} W^*_{2q,m_2}}{[(m_2-m_1)\operatorname{Re} f - i(|m_2|+|m_1|)\operatorname{Im} f + v_2 - v_1](\zeta_{2q} + m_2 \operatorname{Re} f - i|m_2|\operatorname{Im} f - v_1)},$$

$$W_{2q,m} = \frac{[(2q)!]^{\frac{1}{2}}}{(-2)^q q!}\left(\frac{\pi f \kappa}{\pi(1+\kappa)}\right)^{\frac{1}{2}} \sum_{n=\max(0,m)}^{2q}\binom{2q}{n}\binom{-2q-\frac{1}{2}}{n-m}\left(\frac{1-\kappa}{1+\kappa}\right)^{2n-m}, \quad |m_2|+|m_1|\neq 0. \tag{21}$$

The spectrum of combs defined by Eqs. (13) and (14) depends only on the ratio of parameters $\varepsilon$ and $\delta$ which formally can be arbitrarily small. Fig. 4 shows the power of the frequency comb teeth for $\varepsilon/\delta$ =0.1, 0.3, and 0.5 under the assumption that the width of optical resonances, $\operatorname{Im}\zeta_q$, is the same for all of them. It is seen that the bandwidth of combs grows with $\varepsilon/\delta$ and can be significant when this ratio approach unity. At the same time, it follows from the expression for $W_{2q,m}$ in Eq. (21) that the power of teeth slowly vanish as $\varepsilon/\delta \to 1$. The plots in Fig. 4 also show that the behavior of comb tooth heights $\Omega_{2q,m}$ is irregular as a function of quantum numbers $q$ and $m$. This is analogues to the behavior of transition probabilities in a quantum mechanical time-dependent harmonic oscillator described by the same Schrödinger equation [24]. However, for each value of $q$ the comb sub-series with close to flat dependence on $m$ are clearly seen (e.g., squares along brown and light blue curves in Fig. 4(c)). The bandwidth of these series increases proportionally to $q$ and is approximately equal to $1.3q \cdot f$. The behavior of teeth power shown in Fig. 4 is in contrast to their exponential decrease with number $m$ for combs generated by laser pumping [7-11]. Notice that the absence of sharp elastic transmission peaks (for $m = 0$) in the plots of Fig. 4 is due to the fact that the major elastic component of the transmission amplitude has been omitted in Eq. (10).

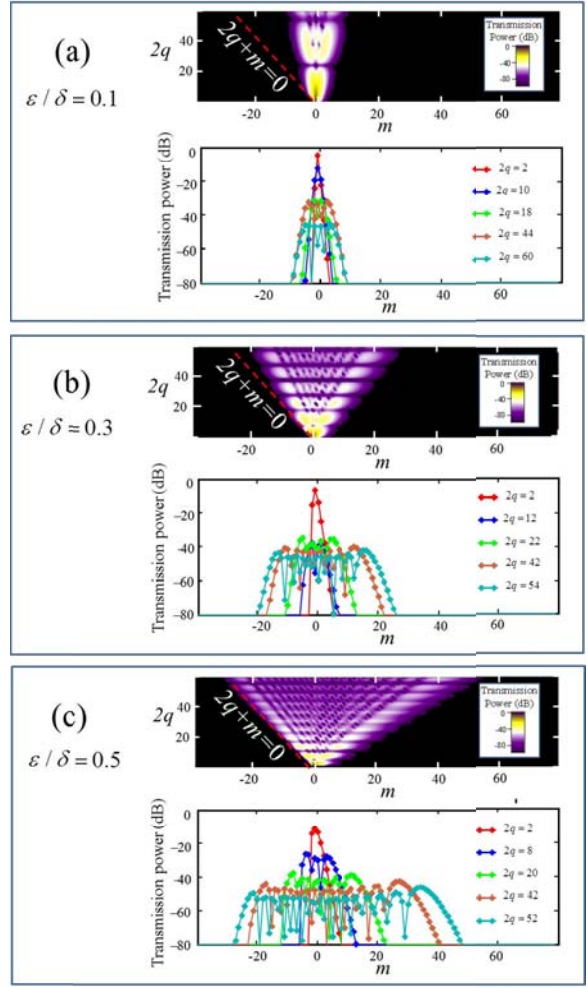

Fig. 4. Surface plots of the power of generated optical frequency combs as a function of quantum numbers $q$ and $m$ and plots of these dependencies for fixed quantum numbers $q$ indicated. (a) – $\varepsilon/\delta$=0.1; (b) – $\varepsilon/\delta$=0.3; (c) – $\varepsilon/\delta$=0.5. Red dashed lines in surface plots indicate the axial ground states of the bottle resonator.

## 6. Lifetime and magnitude of mechanical vibrations and the effect of their nonlinearity

For the eigenfrequency spacing $\Delta \nu_{ax} \sim f \sim 100$ MHz considered, the characteristic $Q_{mech}$ can be as large as $10^4$ [15, 16] corresponding to the lifetime of mechanical vibrations $\tau_{mech} = Q_{mech} / \operatorname{Re} f \sim 0.1$ ms.

Let us estimate the magnitude of mechanical vibrations required for the generation of OFC and the effect of their nonlinearity. The OFC is well defined only if $\text{Im}\,\nu_q \ll \Delta\nu_{ax} \sim f$. For the characteristic eigenfrequency spacing $\Delta\nu_{ax} \sim 100\,\text{MHz}$ and $\nu_q \sim 400\,\text{THz}$, this condition means $Q_{opt} \gg 4\cdot 10^6$. This is realistic for bottle microresonators with $r_0 > 20\,\mu\text{m}$ (see, e.g., [27], where $Q_{opt} = 3\cdot 10^8$ was measured for $r_0 = 19\,\mu\text{m}$). We assume that, due to the smoothness of the bottle resonator radius variation $\Delta r(z)$ and the semiclassical nature of the axial spectrum, the deviation from equidistance of the spectral series considered can be made very small, at least within a relatively small bandwidth. Tuning of eigenfrequencies by $CO_2$ and femtosecond laser post-processing is possible as well [28, 29]. We suggest that the maximum deviation from the equidistance of the spectral series considered, $\max|\delta\nu_m|$ has the order of 1 MHz, similar to [30]. In this case, the exactly solvable model of section 6 can be applied for $\varepsilon \sim \delta \gg 1\,\text{MHz}$. For $r_0 = 20\,\mu\text{m}$ and $\varepsilon \sim \delta \sim 10\,\text{MHz}$, we find the required magnitude of mechanical vibrations $\delta r \sim r_0 \delta / \nu_q \sim 1$ pm. From [20], vibrations with so small magnitude can be excited by the voltage of ~ 0.01 V applied to the piezoelectric actuator described in section 3.

In the absence of noise [31], the frequency of mechanical vibrations $f$ slightly changes during the relaxation time $\tau_{mech}$ due to nonlinear processes. We estimate this change using the one-dimensional nonlinear wave equation which describes the radial deformation of microresonator $u(r)$ [32]:

$$\rho_0 u_{tt} - \beta u_{rr} = \gamma u_{rr} u_r. \tag{22}$$

Here $\rho_0$ is the material density and $\beta$ and $\gamma$ are the second and third order elastic constants. Setting roughly $u_r \sim u/\lambda_s$ and $u_{rr} \sim u/\lambda_s^2$, where $\lambda_s$ is the characteristic wavelength of the excited sound mode of we obtain the ordinary nonlinear differential equation

$$u_{tt} - (2\pi f_0)^2 u = \mu u^2, \quad f_0 = \frac{\beta^{1/2}}{2\pi \rho_0^{1/2} \lambda_s}, \quad \mu = \frac{\gamma}{\rho_0 \lambda_s^3}. \tag{23}$$

Solution of Eq. (23) in the second order in $\mu$ yields the following estimate for the dependence of an eigenfrequency of vibration on its amplitude $\delta r$ [33]:

$$f = f_0(1+\alpha_2), \quad \alpha_2 \sim \left(\frac{\gamma \delta r}{\beta \lambda_s}\right)^2 \tag{24}$$

For silica microresonator, the density $\rho_0 = 2200\,\text{kg/m}^3$ and parameters $\beta \sim 80$ GPa and $\gamma \sim 500$ GPa [34]. Then, for $\delta r = 1$ pm and $\lambda_s \sim 2$ μm, the relative nonlinear variation of comb spacing is $\alpha_2 \sim 10^{-11}$.

## 7. Summary

It is shown that an OFC can be generated mechanically by excitation of an optical bottle microresonator with equidistant spectrum by its natural mechanical vibrations. In practice, small deviations from the spectral equidistance are introduced by fabrication errors. However, these deviations affect the OFC teeth power rather than their equidistance, which is defined by the natural frequency of vibrations $f$. As an example, we determine the OFC generated by a bottle microresonator with parabolic ERV excited by harmonic oscillations. It is found that the power of OFC teeth depends on the ratio of the perturbation of the ERV and deviation of the excitation frequency rather than their actual values. The bandwidth of the generated OFC is equal to $1.3q \cdot f$ where $q$ is the axial quantum number corresponding to the input optical frequency $\nu_q$. Generally, the power of OFCs generated mechanically is inverse proportional to the squared product of their optical and mechanical Q-factors. Provided that these Q-factors are large enough, the power required for the generation of these combs can be remarkably small and only limited by the sensitivity of the optical detectors. A bottle resonator with parabolic ERV is not the only one which possesses equidistant spectral series. Other structures potentially enabling the generation of OFCs mechanically include specially designed bottle resonators with non-parabolic ERV [23] and series of coupled microresonators [35].

## Acknowledgements

The author acknowledges the Royal Society Wolfson Research Merit Award (WM130110) and funding from Horizon 2020 (H2020-EU.1.3.3, 691011). Fruitful discussions with Andrey Matsko are greatly appreciated.